\documentclass[10pt,a4paper]{amsart}

\usepackage{amssymb}
\usepackage{amsmath}
\usepackage{verbatim}
\usepackage{enumerate}
\usepackage{color}
\usepackage{graphicx}
\usepackage{xcolor}
\usepackage{amscd}
\usepackage{amsthm}
\usepackage{epsfig}
\usepackage{eulervm}
\usepackage{subfigure}
\usepackage[
    colorlinks=true,
    linkcolor=black,
    urlcolor=blue,
    citecolor=blue
]{hyperref}
\usepackage{tikz}
\usepackage{braket}
\usepackage[style=alphabetic]{biblatex}
\usepackage{ytableau}
\usetikzlibrary{decorations.pathreplacing}
\usetikzlibrary{calc}

\newcommand{\diag}{\mathsf{diag}}

\theoremstyle{definition}
\newtheorem{remark}{Remark}[section]
\newtheorem{example}[remark]{Example}

\newtheorem{proposition}[remark]{Proposition}
\newtheorem{theorem}[remark]{Theorem}
\newtheorem{definition}[remark]{Definition}
\newtheorem{corollary}[remark]{Corollary}

\begin{document}
%\author{Maxim Gritskov and Savelii Timchenko}
\title[ 
Finite-rank conformal quantum mechanics
]{Finite-rank conformal quantum mechanics}

\begin{abstract}
In this work, we study the simplest example of the landscape of conformal field theories: one-dimensional CFTs with finite-dimensional state space. Following the definition of quantum field theory given by G. Segal in \cite{Segal1988}, we formulate the condition under which a one-dimensional QFT (quantum mechanics) possesses conformal symmetry, and we give a complete classification of conformal Hamiltonians with finite rank. It turns out that correlation functions in such theories are polynomial functions of the underlying geometric data. Moreover, the one-dimensional conformal Ward identities determine their scaling behavior, so that the correlators of the conformal observables are, in fact, homogeneous polynomials.
\end{abstract}

\author{Maxim Gritskov}

\address{Skolkovo Institute of Science and Technology, 121205, Moscow, Russia}
\address{Saint Petersburg State University,
Universitetskaya nab. 7/9, 199034 St. Petersburg, Russia}
\email{m.gritskov@spbu.ru}

\address{Faculty of Physics, National Research University Higher School of Economics, Staraya Basmannaya ul. 21/4s1, 105066, Moscow, Russia}
\email{sgtimchenko@edu.hse.ru}

\author{Saveliy Timchenko}

\maketitle

\setcounter{tocdepth}{3} 
\tableofcontents

\allowdisplaybreaks

\section{Introduction}
One of the most interesting questions in modern quantum field theory is the description of the landscape of conformal theories. Ideologically, all quantum field theories form a manifold, and conformal theories are the zero locus of a distinguished vector field known in physics as the beta function.

To date, many attempts have been made to formalize the concept of quantum field theory \cite{Dedushenko_2023}. The description that comes closest to the concept of space of theories is the functorial description, first proposed by M. Atiyah in \cite{Atiyah1988} for the formalization of topological field theories and later expanded by G. Segal in \cite{Segal1988}. The role of space-time $X$ in this approach is played by $d$-cobordisms, i.e., sheets stretched between two $(d-1)$-manifolds, and quantum field theory is viewed as a functor from the category of geometrically equipped cobordisms to the category of vector spaces. By default, it is assumed that the geometric data is a metric $g$ on space-time $X$. The renormalization group flow on the space of such functors is induced by Weyl transformations on the space of metrics. Theories that are invariant under such a flow are called conformal theories. The partition function of a conformal field theory on a manifold $X$ with metric $g$ depends only on the conformal class of this metric. A systematic pedagogical introduction to Segal's axiomatic approach to quantum field theory can be found in \cite{mnev2025lecturenotesconformalfield, moore2025tasilecturestopologicalfield}. A discussion of the beta function within the framework of the functorial approach is provided in \cite{Gritskov2025}.

Our goal was to consider the simplest example of the landscape of conformal theories. We considered Segal's axioms in dimension $d=1$ and formulated a condition under which a one-dimensional theory is conformal in order to construct an example of a beta function in quantum mechanics. However, it turned out that the condition of scale invariance in dimension $1$ is so strict that, for a fixed dimension of the state space, conformal theories are only a finite number of isolated points in the space of all theories. Thus, conformal quantum mechanics do not have interesting deformations. Despite this, we have discovered that the correlation functions in such theories satisfy a one-dimensional analogue of the conformal Ward identities and possess a number of interesting properties.

The article is structured as follows. In Chapter~\ref{ch:functorial_qm}, we provide a very brief introduction to functional quantum mechanics, define correlation functions, and discuss how the space of quantum mechanical modules is structured. In Chapter~\ref{ch:conformal_qm}, we move on to conformal theories: we define scale-invariant quantum mechanics, study how their space is structured within the landscape of all theories, derive conformal Ward identities for correlation functions, and discuss their consequences.
\section{Functorial quantum mechanics} \label{ch:functorial_qm}
\subsection{Source category}
In the functional picture of $d$-dimensional QFT, we define the source (or the space-time) category $\mathsf{Source}_{d}$ \cite{losev2023tqfthomologicalalgebraelements}. The objects of this category are $(d-1)$-manifolds, and the morphisms are $d$-cobordisms. Morphisms and objects, being manifolds, are usually equipped with consistent geometric structures, such as metrics, connection, etc. Quantum field theory is defined as a symmetric monoidal functor from the source category to $\mathsf{Vect}_{\,\mathbb{C}}$. When $d=1$, the situation is greatly simplified and we can describe many constructions explicitly.
\begin{definition}
Objects of the category $\mathsf{Source}_{1}$ are topological spaces $\mathcal{T}$ consisting of a finite number of points, each of which is assigned an orientation $\pm$:
\begin{equation}
    \mathcal{T}=\{\mathrm{pt}_{1}^{+},...,\mathrm{pt}_{k}^{+},\mathrm{pt}_{k+1}^{-},...,\mathrm{pt}_{n}^{-}\}\,.
\end{equation}
Morphisms between objects $\mathcal{T}_{1}$ and $\mathcal{T}_{2}$ are cobordisms $(X,g)$ equipped with a metric tensor $g\in \Gamma(\mathrm{S}^{2}T^{*}X)$. The cobordism manifold between $\mathcal{T}_{1}$ and $\mathcal{T}_{2}$ is a smooth compact $1$-dimensional $X$ such that $\partial X=\mathcal{T}_{1}\sqcup -\,\mathcal{T}_{2}$, where $-\,\mathcal{T}_{2}$ is a collection of points $\mathcal{T}_{2}$ with opposite orientations. We consider the cobordism $(X,g)$ equivalent to the cobordism $(\tilde{X},\tilde{g})$ if there exists a diffeomorphism $f: X\rightarrow \tilde{X}$ such that $g=f^{*}\tilde{g}$.
\end{definition}

\begin{example}
A line segment of length $\tau$ can be considered as a metric-equipped cobordism between two points (see figure \hyperref[fig:fig1]{1a}).
\end{example}
\begin{example}
    A circle of length $\tau$ can be considered as a metric-equipped cobordism between two empty sets.
\end{example}

\begin{figure}[h]
    \centering
    \subfigure[]{
    \begin{tikzpicture}[baseline, scale=1.5]
      \draw[red, very thick] (0,0) -- (3,0);

      \node[black] at (1.47,-0.2) {$\tau$};
      
      \filldraw[black] (0,0) circle (1.25pt);
      \filldraw[black] (3,0) circle (1.25pt);
    \end{tikzpicture}
    } 
    \hspace{1cm}
    \subfigure[]{
    \begin{tikzpicture}[baseline, scale=1.5]
      \draw[red, very thick] (0,0) -- (3,0);

      \node[black] at (1.47,-0.2) {$\tau_1$};
      
      \filldraw[black] (0,0) circle (1.25pt);
      \filldraw[black] (3,0) circle (1.25pt);
      
      \draw[blue, very thick] (1.5,0.66) circle (0.5);
      
      \node[black] at (2.2,0.66) {$\tau_2$};
    \end{tikzpicture}
    } 
    \caption{}
    \label{fig:fig1}
\end{figure}

\begin{example}
The disjoint union of a line segment with length $\tau_1$ and a circle with length $\tau_2$ is a cobordism between two points (see figure \hyperref[fig:fig1]{1b}).
\end{example}

\begin{figure}[h]
    \centering
    \subfigure[]{
    \begin{tikzpicture}[baseline, scale=1.5]
  \draw[red, very thick] (-1.5,0.5) -- (1.5,0.5);

  \draw[blue, very thick] (-1.5,-0.5) -- (1.5,-0.5);
        \filldraw[black] (-1.5,0.5) circle (1.25pt);
      \filldraw[black] (1.5,0.5) circle (1.25pt);
            \filldraw[black] (-1.5,-0.5) circle (1.25pt);
      \filldraw[black] (1.5,-0.5) circle (1.25pt);

      \node[black] at (-0.03,0.7) {$\tau_1$};
      \node[black] at (-0.03,-0.7) {$\tau_2$};
      
      \node[black] at (-1.5,0.7) {$pt_1$};
      \node[black] at (-1.5,-0.7) {$pt_2$};
      \node[black] at (1.5,0.7) {$pt_1$};
      \node[black] at (1.5,-0.7) {$pt_2$};
\end{tikzpicture}
    } 
    \hspace{1cm}
    \subfigure[]{
    \begin{tikzpicture}[baseline, scale=1.5]
  \draw[red, very thick] (-1.5,0.5) -- (1.5,-0.5);

  \draw[blue, very thick, preaction={draw=white, line width=5pt}] (-1.5,-0.5) -- (1.5,0.5);
        \filldraw[black] (-1.5,0.5) circle (1.25pt);
      \filldraw[black] (1.5,0.5) circle (1.25pt);
            \filldraw[black] (-1.5,-0.5) circle (1.25pt);
      \filldraw[black] (1.5,-0.5) circle (1.25pt);

      \node[black] at (-0.75,0.45) {$\tau_1$};
      \node[black] at (0.75,0.45) {$\tau_2$};
      
      \node[black] at (-1.5,0.7) {$pt_1$};
      \node[black] at (-1.5,-0.7) {$pt_2$};
      \node[black] at (1.5,0.7) {$pt_1$};
      \node[black] at (1.5,-0.7) {$pt_2$};
\end{tikzpicture}
    } 
    \caption{}
    \label{fig:fig2}
\end{figure}
\begin{example}
Two line segments of lengths $\tau_1$ and $\tau_2$ can be considered as two different cobordisms between four points (see figure \hyperref[fig:fig2]{2}).
\end{example}

\begin{proposition}
    Since we are studying monoidal functors, it suffices to consider only connected cobordisms, as any disconnected ones arise from these by applying the monoidal operation (i.e., by taking disjoint unions). Consequently, we need only consider line segments and circles of various lengths, since any other cobordism with the same total length and diffeomorphic to them is regarded as equivalent.
\end{proposition}

\subsection{Partition functions as functors}
\begin{definition}
The quantum mechanical partition function $Z_X$ is an element of the tensor product
\begin{equation}
V_{\Gamma_1}\otimes\dots\otimes V_{\Gamma_m}\otimes V^{*}_{\Gamma_{m+1}}\otimes\dots\otimes V^{*}_{\Gamma_n}\otimes C^{\infty}(\mathbb{R}_{\geq 0}),
\end{equation}
which satisfies the requirement (called cutting axiom) that for any manifold cutting point $\Gamma$, the following relation holds:
\begin{equation}
    \left(Z_{X_1}\otimes Z_{X_2}\right)_{\Gamma}=Z_{X},
\end{equation}
where $(\cdot)_\Gamma$ is a canonical pairing between spaces $V^{*}_{\Gamma}$ and $V_{\Gamma}$. We consider partition functions $Z_1$ and $Z_2$ to be equivalent, if $Z_1=UZ_2 U^{-1}$ for some $U\in \mathsf{GL}(V)$. From now on we do not distinguish equivalent partition functions.
\end{definition}
\begin{remark}
    One sees that this definition of the partition function aligns directly with the definition of a $1$-dimensional QFT. The cutting axiom then follows immediately from functoriality. Moreover, the requirement that partition functions be identified up to conjugation reflects the principle that functors differing only by natural isomorphisms should be regarded as equivalent (conjugation in the category of vector spaces corresponds precisely to such a natural transformation).
\end{remark}
\begin{proposition}
    By choosing the basis, one can always reduce the Hamiltonian to the Jordan form. Thus, the moduli space of quantum mechanics is $\mathsf{End}(V)/\mathsf{GL}(V)$. It consists of $p(n)$ connected components (where $p(n)$ is the number of partitions of $n$). The points in each component are parameterized by the set of eigenvalues.
\end{proposition}
\begin{example}
The partition function $Z_\tau$ on a line segment of length $\tau$ is an element of the tensor product $\mathsf{End}(V)\otimes C^{\infty}(\mathbb{R}_{\geq 0})$, which can be considered as $\mathsf{End}(V)$-valued function on $\mathbb{R}_{\ge0}$. Then the cutting axiom can be written as
\begin{equation}
\label{functoriality}
    Z_{\tau_1}Z_{\tau_2}=Z_{\tau_1+\tau_2},
\end{equation}
and partition function takes form of $Z_{\tau}=e^{-\tau H}$, where $H\in\mathsf{End}(V)$ is an arbitrary operator called Hamiltonian.
\end{example}
\begin{remark}
    Most physically meaningful theories assume that the vector space $V$ is equipped with a non-degenerate scalar product, and the Hamiltonian $H$ is a self-adjoint operator. In this case, $Z_{\tau}$ is the Euclidean evolution operator.
\end{remark}
\begin{example}
The partition function on a circle $\mathrm{S}_\tau$ is just an element of $C^{\infty}(\mathbb{R}_{\geq 0})$. Let's cut the circle at some point $\Gamma$ and get the line segment of length $\tau$ with spaces $V_\Gamma$ and $V^{*}_\Gamma$, associated with its boundaries. Considering the cutting axiom:
\begin{equation}
    Z_{\mathrm{S}_\tau}=\left(Z_{\tau}\right)_{\Gamma}=\mathrm{Tr}_{V_\Gamma}\left(e^{-\tau H}\right).
\end{equation}
\end{example}
\begin{definition} \label{def:correlator}
The correlator of several local observables $O_1, \dots, O_n\in\mathsf{End}(V)$ at points $p_1\ge\dots\ge p_n$ on a circle $\mathrm{S}_\tau$ is given by
\begin{equation} \label{correlation_function}
\braket{O_1(p_1)\dots O_n(p_n)}_{\mathrm{S}_\tau}= \mathrm{Tr}_V\left(e^{-(\tau-\tau_{n1}) H}O_1 e^{-\tau_{21}H}O_2\dots e^{-\tau_{nn-1}H}O_n\right),
\end{equation}
where $\tau_{nm}$ is the length of the circle segment between points $p_n$ and $p_m$. Note that the ordering of the points $p_1,\dots,p_n$ is defined on the line segment before it is glued (see figure \hyperref[fig:circle_gluing]{3}) into a circle; the same applies to the lengths $\tau_{nm}$.
\end{definition}
\begin{figure}[h!] \label{fig:circle_gluing}
\centering

\begin{tikzpicture}[
    point/.style={circle, fill=black, inner sep=1.5pt},
    label text/.style={font=\small}, scale=0.9
]

    %% --- LEFT SIDE: LINEAR SEGMENT ---
    \coordinate (start) at (-0.5,0);
    \coordinate (L_pn)   at (0,0);
    \coordinate (L_pnm1) at (1.4,0);
    \coordinate (L_p2)   at (2.4,0); 
    \coordinate (L_p1)   at (3.8,0);
    \coordinate (end) at (4.3,0);

    \draw[very thick, blue] (start) -- (L_pn);
    \draw[very thick, blue] (L_pn) -- (L_pnm1);
    \draw[very thick, dotted] (L_pnm1) -- (L_p2);
    \draw[very thick, blue] (L_p2) -- (L_p1);
    \draw[very thick, blue] (L_p1) -- (end);

    \node[point, label=below:$p_n$] at (L_pn) {};
    \node[point, label=below:$p_{n-1}$] at (L_pnm1) {};
    \node[point, label=below:$p_2$] at (L_p2) {};
    \node[point, label=below:$p_1$] at (L_p1) {};

    \node[above] at ($(L_pn)!0.5!(L_pnm1)$) {$\tau_{nn-1}$};
    \node[above] at ($(L_p2)!0.5!(L_p1)$) {$\tau_{21}$};

    \draw[decorate, decoration={brace, amplitude=8pt, raise=15pt}, thick] 
        (start) -- (end) node[midway, above=25pt] {$\tau$};

    %% --- CENTER: ARROW ---
    \draw[->, line width=1pt] (5.0, 0) -- (6.5, 0);

    %% --- RIGHT SIDE: CIRCLE ---
    \begin{scope}[shift={(10,0)}]
        \def\R{1.8}
        
        \draw[very thick, blue] (135:\R) arc (135:225:\R);
        \node at (180:\R+0.6) {$\tau_{nn-1}$};

        \draw[very thick, dotted] (225:\R) arc (225:315:\R);

        \draw[very thick, blue] (315:\R) arc (315:405:\R);
        \node at (0:\R+0.4) {$\tau_{21}$};

        \draw[very thick, blue] (45:\R) arc (45:135:\R);
        \node at (90:\R+0.3) {$\tau - \tau_{n1}$};

        \node[point] at (135:\R) {};
        \node[font=\small] at (135:\R+0.35) {$p_n$};

        \node[point] at (225:\R) {};
        \node[font=\small] at (225:\R+0.35) {$p_{n-1}$};

        \node[point] at (315:\R) {};
        \node[font=\small] at (315:\R+0.35) {$p_2$};

        \node[point] at (45:\R) {};
        \node[font=\small] at (45:\R+0.35) {$p_1$};
        
    \end{scope}

\end{tikzpicture}
\caption{}
\end{figure}

\begin{remark}
    Definition~\ref{def:correlator} is not the most general. In the functorial QFT, a different definition is more commonly used, as described for example in \cite{kontsevich2021wickrotationpositivityenergy}.
\end{remark}

\begin{example}
The correlator of two observables $O_1,O_2$ on a circle $\mathrm{S}_\tau$ equals to
\begin{equation} \label{two_point_correlator}
\braket{O_1(p_1)O_2(p_2)}_{\mathrm{S}_\tau}=\mathrm{Tr}_V\left(e^{-(\tau-\tau_{21}) H}O_1e^{-\tau_{21}H}O_2\right).
\end{equation}
\end{example}

\section{Conformal quantum mechanics} \label{ch:conformal_qm}
\subsection{Partition functions of conformal theories}
Now we will give a motivation for the definition of conformal quantum mechanics. The partition function $Z_{\tau}$ is called Weyl invariant if the space $V$ is a representation of the Lie algebra $\mathcal{X}(\mathbb{R}_{\geq 0})$ of vector fields in the space of geometric data and the following identity is held:
\begin{equation}
\label{Weyl}
    \mathcal{L}_{v}Z_{\tau}=[L_{v},Z_{\tau}]\,.
\end{equation}
Here, $\mathcal{L}_{v}$ is the derivative along the vector field $v\in\mathcal{X}(\mathbb{R}_{\geq 0})$ and $L_{v}$ is the generator of the representation corresponding to the vector field $v$. It is easy to show that there are no non-trivial solutions satisfying condition \eqref{functoriality} for such an equation. This is because if we consider \eqref{Weyl} for $v_{n}=t^{n}\partial_{t}$, where $n\geq 1$, then the system will be inconsistent for $n\geq 2$. Therefore, Weyl-invariant quantum mechanics does not exist. However, if we consider only the $n=1$ case, then solutions to such an equation exist, and we will call them scale-invariant or conformal quantum mechanics.
\begin{definition} \label{def:conformal_qm}
We will call a theory with partition function $Z_{\tau}$ conformal quantum mechanics if for any $\Lambda>0$ there exists $U(\Lambda)\in\mathsf{GL}(V)$ such that
\begin{equation} \label{conformal_definition}
    Z_{\Lambda \tau}=U(\Lambda)Z_{\tau}U^{-1}(\Lambda)
\end{equation}
\end{definition}
\begin{remark}
     For the conformal partition function on a circle $\mathrm{S}_\tau$ we get a known definition of the conformal partition function $Z_{\Lambda \tau}=Z_\tau$, whose two-dimensional counterpart was first given by A. Zamolodchikov and Al. Zamolodchikov in \cite{zamolodchikov1989conformal}.
\end{remark}
\begin{definition}
We define the dilatation generator $L\in\mathsf{End}(V)$ such that
\begin{equation} \label{dilatation_generator}
    U(\Lambda)=\Lambda^{-L}.
\end{equation}
\end{definition}
\begin{proposition}
    There is the following commutation relation
    \begin{equation} \label{dilatation_commutation}
        [L,H]=-H.
    \end{equation}
    The operators $L$ and $H$ therefore generate the Lie algebra $\mathfrak{aff}(1)^{\mathbb{C}}$. This algebra is equivalently known as the Borel subalgebra of $\mathfrak{sl}(2,\mathbb{C})$.\\
    \textbf{Proof}.
    This commutation relation is equivalent to $\mathsf{ad}_{-L} H=H$, which in turn is equivalent to equation \eqref{conformal_definition}.
    \qed
\end{proposition}
\begin{remark}
    For the same Hamiltonian $H$, there could exist several non-equivalent dilatation generators $L$ such that equation \eqref{dilatation_commutation} is satisfied.
\end{remark}
\begin{theorem} \label{thm:zero_spectrum}
The conformal Hamiltonian of finite rank has a zero spectrum.\\
\textbf{Proof}. By substituting $Z_{\tau}=e^{-\tau H}$ into (\ref{conformal_definition}), we get
\begin{equation} \label{hamiltonian_dilatation}
    \Lambda H=U(\Lambda)H U^{-1}(\Lambda).
\end{equation}
On the one hand,
\begin{equation}
    \mathrm{Spec}(H)=\mathrm{Spec}(UHU^{-1}),
\end{equation}
and, on the other hand
\begin{equation}
    \mathrm{Spec}(\Lambda H)=\{\Lambda\sigma:\sigma\in\mathrm{Spec}(H)\}.
\end{equation}
Then we conclude that the only possible way for (\ref{conformal_definition}) to be true for any $\Lambda\in\mathbb{R}_{\ge0}$ is for the spectrum of Hamiltonian to be zero.\qed
\end{theorem}
\begin{corollary} \label{crl:polynomials}
Correlation functions are polynomials in $\tau_{21}, \tau_{32},\dots, \allowbreak \tau_{nn-1},\tau$.\\
\textbf{Proof}. Because we identify Hamiltonians up to their conjugacy class, we can always choose the Hamiltonian to be a Jordan matrix with a zero spectrum. As we know, those matrices are nilpotent, so the partition function $Z_{\tau}=e^{-\tau H}$ is a polynomial function in $\tau$, and correlators (\ref{correlation_function}) become polynomials in $\tau_{21},\tau_{32},\dots,\tau_{nn-1}, \tau$.\qed
\end{corollary}
\begin{corollary}
Conformal Hamiltonians are classified by Young diagrams.\\
\textbf{Proof}. As we noted in Corollary~\ref{crl:polynomials}, we can always choose the Hamiltonian to be a Jordan matrix with a zero spectrum and an arbitrary order of the Jordan blocks. That gives us an opportunity to construct the unique Young diagram, where each row corresponds to one Jordan block of the Hamiltonian, and the length of that row coincides with the dimension of the corresponding Jordan block.\qed
\end{corollary}
\begin{example} \label{ex:young}
We can identify the following Hamiltonians and Young diagrams:
\begin{equation}
\begin{pmatrix}
    0 & 1 & 0 & 0 \\
    0 & 0 & 0 & 0\\
    0 & 0 & 0 & 1\\
    0 & 0 & 0 & 0
\end{pmatrix}
\ \longleftrightarrow\ \ \scalebox{0.8}{\raisebox{0.32em}{$\ydiagram{2,2}$}}
\ ;\ \ \ \ 
\begin{pmatrix}
    0 & 1 & 0 & 0 \\
    0 & 0 & 0 & 0\\
    0 & 0 & 0 & 0\\
    0 & 0 & 0 & 0
\end{pmatrix}
\ \longleftrightarrow\ \ \scalebox{0.8}{\raisebox{1.05em}{$\ydiagram{2,1,1}$}}\ .
\end{equation}
\end{example}
\begin{proposition}
    In each connected component of the moduli space of quantum mechanics $\mathsf{End}(V)/\mathsf{GL}(V)$, there is exactly one point corresponding to a conformal theory: it is a Hamiltonian with all eigenvalues equal to zero.
\end{proposition}
\begin{corollary}
    Since the space of conformal theories is a finite number of points, there are no non-trivial deformations of finite-rank conformal quantum mechanics. 
\end{corollary}

\begin{remark}
    Our primary focus will be on diagonalizable $L$, since their diagonalizability ensures that the space of local observables splits into a direct sum of conformal multiplets, as will be established in Corollary~\ref{crl:conformal_multiplet}.
\end{remark}

\begin{theorem} \label{thm:ad_L_diagonalization}
    If the operator $L$ is diagonalizable, then its adjoint action $\mathsf{ad}_{L}$ is diagonalizable as well. Moreover, the eigenvalues of $\mathsf{ad}_{L}$ are precisely the differences $\Delta_{ij}=\sigma_i-\sigma_j$, where $\sigma_i$ and $\sigma_j$ range over the eigenvalues of $L$.\\
    \textbf{Proof}. Assume that $L$ is diagonalizable. Then there exists a basis $e_i$ of $V$ consisting of eigenvectors of $L$ with eigenvalues $\sigma_i$:
    \begin{equation}
        L e_i=\sigma_i e_i.
    \end{equation}
    For each pair of indices $i,j$, let $E_{ij}\in\mathsf{End}(\mathcal{V})$ denote the elementary endomorphism sending $e_j$ to $e_i$ and all other basis vectors to $0$. Let's directly compute $[L, E_{ij}]$ by applying it to $e_k$. Only $k=j$ gives nonzero result, and
    \begin{equation}
        [L,E_{ij}]e_j=L e_i-\sigma_j E_{ij}e_j=(\sigma_i-\sigma_j)e_i=(\sigma_i-\sigma_j)E_{ij}e_j.
    \end{equation}
    Thus,
    \begin{equation}
        [L,E_{ij}]=(\sigma_i-\sigma_j)E_{ij}.
    \end{equation}
    Therefore each $E_{ij}$ is an eigenvector of the adjoint operator $\mathsf{ad}_{L}$ with eigenvalue $\Delta_{ij}=\sigma_i-\sigma_j$, which we'll call the conformal dimension of $E_{ij}$. The set $E_{ij}$ forms a basis of $\mathsf{End}(V)$, hence $\mathsf{ad}_{L}$ is diagonalizable with spectrum $\Delta_{ij}=\sigma_i-\sigma_j$.\qed
\end{theorem}

\begin{corollary} \label{crl:conformal_multiplet}
    When $\mathsf{ad}_{L}$ is diagonalizable, the space of observables $\mathsf{End}(V)$ splits into a direct sum of $\mathsf{ad}_{L}$-eigenspaces, which we call spaces of conformal observables:
    \begin{equation}
        \mathsf{End}(V)=\bigoplus_{\Delta\in\mathsf{Spec}(\mathsf{ad}_{L})}\mathsf{Conf}_{\Delta}(V)
    \end{equation}
\end{corollary}
\begin{definition}
   A local observable $O$ is called conformal with dimension $\Delta$ if $O$ is an eigenvector of $\mathsf{ad}_{L}$ with eigenvalue $\Delta$, or, equivalently, if $O\in\mathsf{Conf}_{\Delta}(V)$.
\end{definition}

\begin{example}
    The operator $L$ admits the following explicit form; here, the partition $\lambda$ is the one associated with the Hamiltonian:
    \begin{equation}
        L=\bigoplus_{i=1}^{l(\lambda)}\diag\{0,1,\dots,\lambda_i-1\}.
    \end{equation}
    The quantity $l(\lambda)$ denotes the number of rows in the Young diagram associated with the partition $\lambda$, and $\lambda_i$ is the length of its $i$-th row. It suffices to verify equation (\ref{dilatation_commutation}) block by block. Let $L_{i}=\diag\{0,1,\dots,\lambda_i-1\}$ and let $H_i$ be a Jordan block of size $\lambda_i$. Denote by $e_k$ the eigenbasis of $L_{i}$, defined via $H_i e_1 = 0$ and $H_i e_k = e_{k-1}$ for $k\ge 2$. Then $L_{i} e_k = (k-1)e_k$ for all $k$. For $k=1$ we have
    \begin{equation}
        [L_{i},H_i]e_1 = 0 = -H_i e_1.
    \end{equation}
    For $k\ge 2$,
    \begin{equation}
        [L_{i},H_i]e_k = L_{i} e_{k-1}-(k-1)H_ie_k=-e_{k-1}=-H_i e_k.
    \end{equation}
    Because the equality is satisfied on the basis $e_k$ for each Jordan block $H_i$, we conclude that the identity (\ref{dilatation_commutation}) holds in general. For such $L$, the endomorphisms $E_{ij}$ become the standard matrix units. Then the spectrum of $\mathsf{ad}_{L}$ takes the form
    \begin{equation}
        \Delta_{ij}=c(i)-c(j),
    \end{equation}
    where $c(i)$ is the column index of the $i$-th box in the Young diagram $\lambda$.
\end{example}

\subsection{Conformal Ward identities}
\begin{proposition}
In conformal QM, for any correlation functions of observables on a circle $\mathrm{S}_\tau$, the following identity, called the Ward identity, can be written:
\begin{equation} \label{ward_identity}
    \braket{O_1(\Lambda p_1)\dots O_n(\Lambda p_n)}_{\mathrm{S}_{\Lambda \tau}}=\braket{\left(\Lambda^{\mathsf{ad}_{L}}O_1\right)(p_1)\dots\left(\Lambda^{\mathsf{ad}_{L}}O_n\right)(p_n)}_{\mathrm{S}_\tau}.
\end{equation}
Scaling points on a circle refers to scaling the lengths of the segments between them.\\
\textbf{Proof.} The identity follows directly from Definition~\ref{def:correlator}:
\begin{eqnarray}
    \braket{O_1(\Lambda p_1)\dots O_n(\Lambda p_n)}_{\mathrm{S}_{\Lambda \tau}}=\\
    \notag=\mathrm{Tr}_V\left(e^{-\Lambda (\tau-\tau_{n1}) H}O_1e^{-\Lambda\tau_{21}H}O_2\dots e^{-\Lambda \tau_{nn-1}H}O_n\right)=\\
    \notag=\mathrm{Tr}_V\left(e^{-(\tau-\tau_{n1}) H}\Lambda^{L} O_1 \Lambda^{-L}\dots e^{-\tau_{nn-1}H}\Lambda^{L}O_n \Lambda^{-L}\right)=\\
    \notag=\braket{\left(\Lambda^{\mathsf{ad}_{L}}O_1\right)(p_1)\dots\left(\Lambda^{\mathsf{ad}_{L}}O_n\right)(p_n)}_{\mathrm{S}_\tau}.
\end{eqnarray}\qed
\end{proposition}
\begin{proposition}
The Ward identity (\ref{ward_identity}) for $n$ conformal observables $O_1,\dots,O_n$ with corresponding conformal dimensions $\Delta_1,\dots,\Delta_n$ can be written as
\begin{equation}
    \braket{O_1(\Lambda p_1)\dots O_n(\Lambda p_n)}_{\mathrm{S}_{\Lambda \tau}}=\Lambda^{\Delta_1+\dots+\Delta_n}\braket{O_1(p_1)\dots O_n(p_n)}_{\mathrm{S}_\tau}.
\end{equation}
\end{proposition}
\begin{corollary}
Correlation functions of conformal observables in conformal QM are homogeneous polynomials in $\tau_{21},\tau_{32},\dots,\tau_{nn-1}, \tau$.
\end{corollary}
\begin{example}
    The Hamiltonian is a conformal observable with conformal dimension $\Delta=-1$, as can be seen directly from Definition~\ref{def:conformal_qm}.
\end{example}
\begin{example}
    For two conformal observables the following expression holds:
    \begin{equation}
        \braket{O_1(\Lambda p_1)O_2(\Lambda p_2)}_{\mathrm{S}_{\Lambda\tau}}=\Lambda^{\Delta_1+\Delta_2}\braket{O_1(p_1)O_2(p_2)}_{\mathrm{S}_{\tau}}.
    \end{equation}
    By expanding the exponents in (\ref{two_point_correlator}), we get
    \begin{equation}
        \braket{O_1(p_1)O_2(p_2)}_{\mathrm{S}_\tau}=\sum_{k,m=0}^{N-1}\frac{(\tau_{21}-\tau)^k\cdot(-\tau_{21})^m}{k!\cdot m!}\mathrm{Tr}_V\left(H^k O_1 H^m O_2\right),
    \end{equation}
    where $N$ is the Hamiltonian nilpotency index. Since we want the two-point correlator of two conformal observables to be a homogeneous polynomial in $\tau_{21},\tau$, the only way to ensure this is to impose the constraint $k+m=\Delta_1+\Delta_2=\delta$:
    \begin{equation}
        \braket{O_1(p_1)O_2(p_2)}_{\mathrm{S}_\tau}=\sum_{k=0}^{N-1}\frac{(\tau_{21}-\tau)^k\cdot(-\tau_{21})^{\delta-k}}{k!\cdot (\delta-k)!}\mathrm{Tr}_V(H^k O_1 H^{\delta-k} O_2).
    \end{equation}
\end{example}
\begin{corollary}
    The correlation function of $n$ conformal observables $O_1,\dots,O_n$ with negative or non-integer total conformal dimension $\Delta_1+\dots+\Delta_n$ is equal to zero. Otherwise, the correlation function would scale as a negative or non-integer power, which is impossible, since it must be a polynomial.
\end{corollary}

\begin{remark}
    Correlation functions of conformal observables with $\Delta=0$ are constant. Local observables with this property are called topological. Moreover, conformal observables with $\Delta=0$ form an associative algebra, because $\mathsf{ad}_{L}$ is a derivation. Note that, in contrast to the algebra of $\Delta=0$ conformal observables in $2$-dimensional CFT, the corresponding algebra in conformal QM is not commutative.
\end{remark}

\section{Conclusion}
In this work, we explored the space of finite-rank conformal quantum mechanics. Assuming that the dilatation generator $L$ is diagonalizable, we classified all corresponding conformal theories and considered the conformal Ward identities, which led to interesting constraints on the correlation functions.

A natural next step is to extend this analysis to representations in which the dilatation generator $L$ is not diagonalizable but instead takes Jordan form. These representations are the one-dimensional analogue of logarithmic conformal field theories \cite{Cardy_2013}. In such theories, the space of local observables no longer splits into a direct sum of $\mathsf{ad}_{L}$-eigenspaces, making it especially interesting to understand how the one-dimensional Ward identities constrain correlation functions in logarithmic conformal quantum mechanics.

\section*{Acknowledgements}
We are grateful to Andrey Losev and Kirill Matirko for helpful discussions. We are grateful to the Shanghai Institute for Mathematics and Interdisciplinary Sciences for providing us a workspace. The first author is supported by the Ministry of Science and Higher Education of the Russian Federation (Agreement No. 075-15-2025-013).

\printbibliography

\end{document}